\documentclass{article} % For LaTeX2e
\usepackage[final]{colm2025_conference}

\usepackage{microtype}
\usepackage{hyperref}
\usepackage{url}
\usepackage{booktabs}
\usepackage{graphicx}
\usepackage{float}

\usepackage{lineno}

\definecolor{darkblue}{rgb}{0, 0, 0.5}
\hypersetup{colorlinks=true, citecolor=darkblue, linkcolor=darkblue, urlcolor=darkblue}

\title{Wartime Media Dynamics in Emerging Democracies: \\ Case Study of Pakistani Media in May 2025 Indo-Pak Conflict}

\author{Taaha S.~Bajwa \\
Independent\\
\texttt{taaha.s.bajwa@gmail.com} \\
}

\begin{document}

\ifcolmsubmission
\linenumbers
\fi

\maketitle

\begin{abstract}
% Democracies fundamentally rely on voices of opposition and social dissent to function effectively. However, in emerging democracies, freedom of speech is often restricted, and the will of the people is frequently threatened by non-democratic forces, resulting in limited media coverage of anti-government voices. Major geopolitical events, such as regional conflicts, can further contribute to the suppression of dissenting views.
% This study aimed to conduct an exploratory data analysis of Pakistani media coverage of national news before, during, and after the India-Pakistan conflict of May 2025. Approximately 2,600 national news articles from three mainstream Pakistani newspapers were analyzed. A large language model (LLM) agent was employed to classify news articles into categories such as war-related coverage and social dissent. The analysis revealed that media coverage of the conflict significantly overshadowed and suppressed reporting related to political opposition and dissent.
% These findings suggest that regional conflicts can have a detrimental impact on democratic processes and freedom of speech, underscoring the need for stronger protections of press independence in conflict-prone regions.

Democracies rely on opposition and dissent to function, but in emerging democracies, freedom of speech is often restricted—an effect that intensifies during regional conflicts. This study examines how the India-Pakistan conflict of May 2025 influenced Pakistani media coverage. Analyzing approximately 2,600 news articles from three major newspapers using a large language model (LLM), the study found that war-related reporting significantly overshadowed coverage of political opposition and dissent. These findings highlight how conflict can marginalize democratic discourse, reinforcing the need to safeguard press freedom in volatile regions.
\end{abstract}

\section{Introduction}
Democracies depend on opposition voices and public dissent to remain healthy, but in transitional democracies, free expression is often limited, restricting media coverage of anti-government perspectives. This suppression intensifies during regional conflicts, as media outlets tend to align with official narratives \citep{framingconflict}, leaving little space for critical reporting. These conditions highlight the need for systematic analysis of how war impacts media coverage.

\section{Objectives}
This study aims to analyze the impact of war on national media coverage of dissent. Specifically, it investigates how coverage of political opposition is affected—particularly whether such voices are overshadowed or suppressed—during times of armed conflict. By examining this relationship, the research seeks to enhance our understanding of how regional wars can undermine democratic processes and shift media focus away from domestic political discourse.

\section{Methodology}
\subsection{Dataset}
A total of 2,639 news articles were collected from the national news sections of three major English-language newspapers in Pakistan. The dataset covers a two-month period from April 1, 2025, to May 31, 2025, encompassing the time before, during, and after the India-Pakistan conflict. Only articles published under the national news category were included in the analysis to ensure relevance to domestic political and conflict-related coverage. All data was collected from publicly accessible web pages, and no personal or subscriber information was used.

The two-month timeframe was divided into four distinct periods for analysis. The first period, referred to as the pre-tension phase, spans from April 1 to April 22. The second period, the tension phase, covers April 22 to May 7. This phase begins with the Pahalgam terror attack in the Indian Occupied Kashmir on April 22, which escalated tensions as India accused Pakistan of supporting the militant group involved. The third period, the war phase, extends from May 7 to May 10, during which both countries launched aerial and missile strikes on each other’s territory. Finally, the fourth period, the post-war phase, runs from May 10 to May 31, following a ceasefire agreement on May 10.

\subsection{Topic Modeling}
To accurately determine whether a news article focused on war or political opposition, traditional topic modeling techniques such as Latent Dirichlet Allocation (LDA) \cite{Blei_LDA} and class-based TF-IDF approaches like BERTopic \cite{grootendorst2022bertopic} proved inadequate. This limitation arises from the inherently multifaceted nature of news reporting, where a single article often covers multiple overlapping themes.

For instance, an article discussing how a neighboring country reduced water flow in the Chenab River—resulting in drought concerns—might be misclassified under environmental issues, despite repeatedly framing the event as an act of aggression. Likewise, an article covering the military trials of political prisoners, which is fundamentally about political dissent, might be misidentified as war-related due to the frequent use of terms like "military", "mutiny", "national security threats" and "propaganda weakening the state."

To address this challenge, a large language model (LLM) agent was used to go over each article individually and classify it into one of three categories: \textbf{India Pakistan Conflict}, \textbf{Opposition and Dissent}, or \textbf{Other}. The classification was performed using GPT-4.1 Nano \cite{openai2025gpt41}. The prompt used for this task is provided in the Appendix \ref{table:llm_prompt}.

\subsection{Clustering}
For clustering, all articles were first converted into embeddings using OpenAI’s text-embedding-3-small model \citep{openai2024textembeddings}. To visualize the clustering, t-SNE was applied to reduce the high-dimensional embeddings to two dimensions, using hyperparameters mentioned in Appendix \ref{cluster-params}. The resulting 2D projection was used to plot the articles, which were color-coded according to their assigned cluster labels.

\section{Key Findings}
\begin{figure}[t]
\begin{center}
  \includegraphics[width=0.9\linewidth]{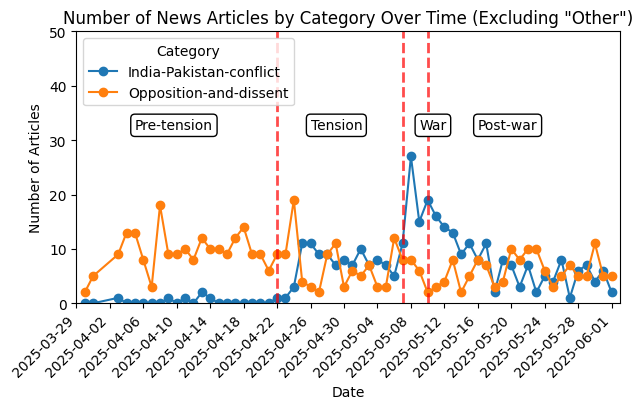}
\end{center}
\caption{Media coverage trends during Pre-tension, Tension, War and Post-war phase.}
\label{fig:coverage_time_series}
\end{figure}

\begin{figure}[H]
\begin{center}
  \includegraphics[width=0.9\linewidth]{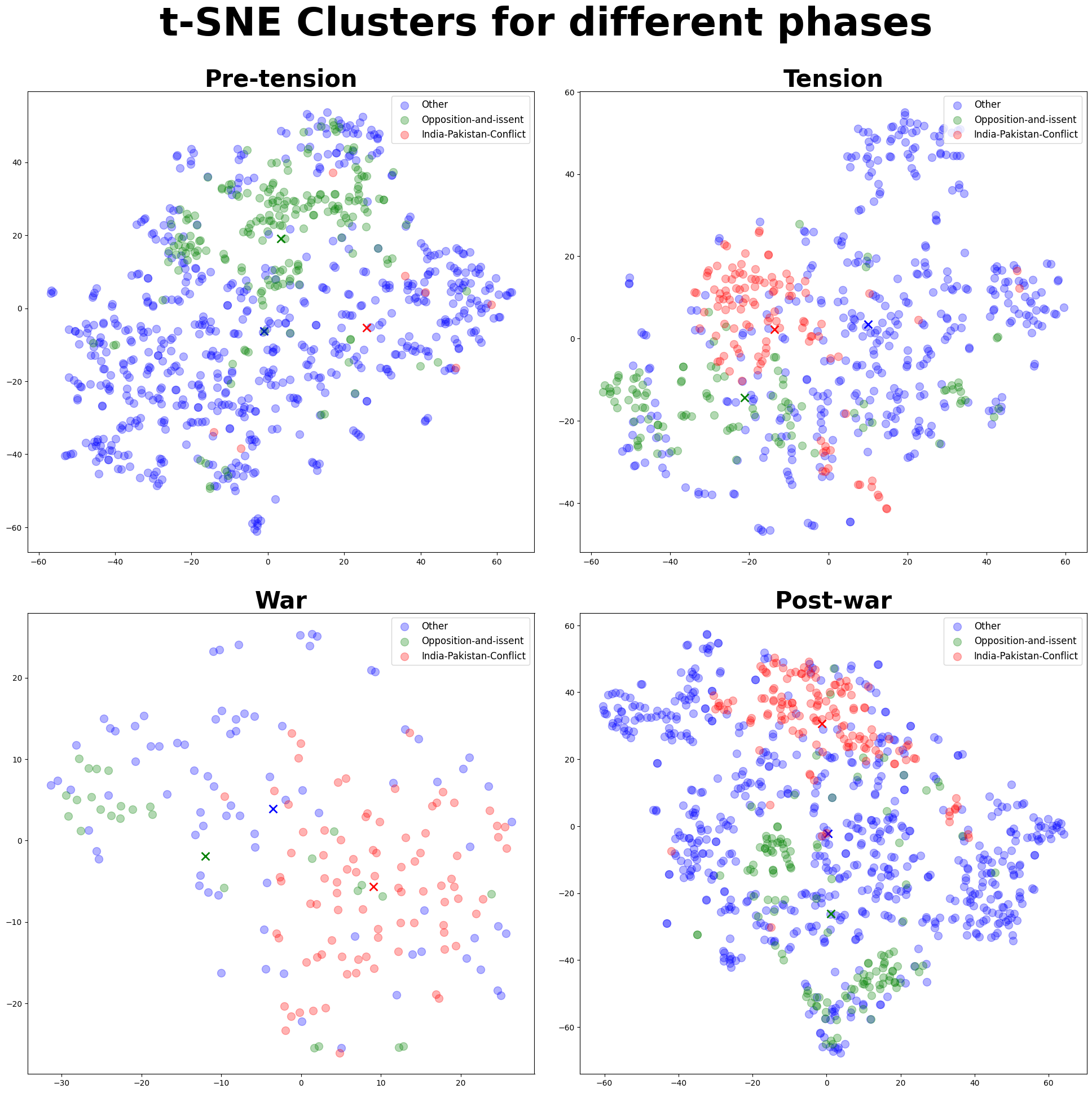}
\end{center}
\caption{t-SNE clusters for Pre-tension, Tension, War and Post-war phase.}
\label{fig:cluster_visualization}
\end{figure}

As shown in Fig. \ref{fig:coverage_time_series} and Fig. \ref{fig:cluster_visualization}, during the pre-tension phase, media coverage of opposition and dissent was relatively high. As tensions began to rise during the tension phase, conflict-related coverage increased and gradually overtook dissent-focused reporting. In the war phase, corresponding to the days of active warfare, coverage of the conflict spiked sharply, while opposition-related reporting dropped significantly. In the post-war phase, although media attention to the conflict began to decline, it continued to receive considerable coverage, whereas opposition reporting remained suppressed. This trend suggests that once the conflict narrative takes hold, it not only dominates media space during the war but also continues to overshadow democratic discourse even after active hostilities have ended.

\section{Future Directions}
In the future, this analysis can be extended by conducting a comparative study of Indian media coverage during the same conflict period to examine whether similar patterns of dissent suppression emerged across the border. Such a comparison would offer deeper insights into how nationalistic narratives shape media behavior in both countries during times of conflict. Additionally, analyzing social media platforms could provide a complementary perspective—revealing whether user-generated content mirrors the suppression seen in traditional media, or if social media serves as an alternative space for dissent when conventional outlets are constrained during wartime.

\bibliography{colm2025_conference}
\bibliographystyle{colm2025_conference}

\appendix
\section{Appendix}
\subsection{Prompt}
Below is the prompt used by LLM agent to classify each news article.
% \begin{table}[H]
% \centering
% \begin{tabular}{l}
% \toprule
% \textbf{Prompt} \\
% \midrule
% You are a news classifier.
% You are given a news article and you need to classify it into one of the following categories:
% - India-Pakistan-conflict
% - Opposition-Parties-and-dissent
% - Other

% If it is related at anything related to India Pakistan conflict (and its effects on Pakistan and India), then classify it as India-Pakistan-conflict.
% If it is related at anything related to opposition and dissent(This include all news of PTI and its trials which is oppostion party of Pakistan, any kinds of protests against goivernment by any segment of society or other parties),
% then classify it as Opposition-Parties-and-dissent.
% If it doesn't fall into any of the above categories, then classify it as Other.

% The news article is:
% \bottomrule
% \end{tabular}
% \caption{Prompt used for LLM agent}
% \label{table:neuron_parts}
% \end{table}

\begin{table}[H]
\centering
\begin{tabular}{p{0.9\linewidth}}  % Wraps text at 90% of text width
\toprule
\textbf{Prompt} \\
\midrule
You are a news classifier. You are given a news article and you need to classify it into one of the following categories:
\begin{itemize}
  \item India-Pakistan-conflict
  \item Opposition-and-dissent
  \item Other
\end{itemize}

If it is related to anything concerning the India-Pakistan conflict (and its effects on Pakistan and India), then classify it as India-Pakistan-conflict.  
If it is related to anything concerning opposition and dissent (This includes all news about PTI and its trials—PTI being the opposition party in Pakistan—as well as any protests against the government by any segment of society or other parties), then classify it as Opposition-and-dissent.  
If it doesn't fall into any of the above categories, then classify it as Other.

The news article is: \\
\bottomrule
\end{tabular}
\caption{Prompt used for LLM agent}
\label{table:llm_prompt}
\end{table}

\begin{table}[H]
\centering
\begin{tabular}{p{0.5\linewidth}}
\toprule
\textbf{Structured Output Class (Pydantic)} \\
\midrule
\centering
\ttfamily
\begin{tabular}{@{}l@{}}
\\
class NewsClassifier(BaseModel): \\
\hspace{4mm}news\_type: Literal[ \\
\hspace{8mm}"India-Pakistan-conflict", \\
\hspace{8mm}"Opposition-and-dissent", \\
\hspace{8mm}"Other" \\
\hspace{4mm}] \\
\end{tabular}
% \bottomrule
\end{tabular}
\caption{Structured output format returned by the LLM agent}
\label{table:pydantic_output}
\end{table}

\subsection{t-SNE hyperparameters}
\begin{table}[H]
\begin{center}
\begin{tabular}{ll}
\toprule
\multicolumn{1}{c}{\bf Hyperparameter}  &\multicolumn{1}{c}{\bf Value} \\
\midrule
Number of Components         &2 \\
Perplexity             &15 \\
Learning Rate             &200 \\
\bottomrule
\end{tabular}
\end{center}
\caption{t-SNE hyperparameters for clustering}\label{cluster-params}
\end{table}

\end{document}